\documentclass[aps,prd,preprintnumbers,showpacs,showkeys,nofootinbib,superscriptaddress,fleqn,floatfix,tightenlines,10pt]{revtex4-1}
\usepackage{amsmath,amsfonts,amssymb,amscd,amsxtra,amsthm}
\usepackage{graphicx}  
\usepackage{epstopdf}
\usepackage{dcolumn}  
\usepackage{bm}          
\usepackage{slashed}
\usepackage[utf8]{inputenc}
\usepackage{CJK}
\usepackage{cancel}
\usepackage{mathtools}
\usepackage{amsbsy}
\usepackage{amstext}
\usepackage[normalem]{ulem} 
\usepackage[dvipsnames]{xcolor} 
\usepackage{array}
\usepackage{slashed}
\renewcommand\sout{\bgroup \color{red} \ULdepth=-.5ex \ULset}

\usepackage[caption = false]{subfig}
\usepackage{multirow}

\begin{document}
\preprint{INHA-NTG-13/2018}
%
\title{In-medium properties of SU(3) baryons}
%
%
\author{Ki-Hoon Hong}
\email{kihoon@inha.edu}
\affiliation{Department of Physics, Inha University, Incheon 22212,
 Republic of  Korea}
\author{Ulugbek Yakhshiev}
\email{yakhshiev@inha.ac.kr}
\affiliation{Department of Physics, Inha University, Incheon 22212,
 Republic of  Korea}
\author{Hyun-Chul Kim}
\email{hchkim@inha.ac.kr}
\affiliation{Department of Physics, Inha University, Incheon 22212,
 Republic of  Korea}
\affiliation{
Advanced Science Research Center, Japan Atomic Energy
  Agency, Shirakata, Tokai, Ibaraki, 319-1195, Japan}
\affiliation{School of Physics, Korea Institute for Advanced Study
 (KIAS), Seoul 02455, Republic of Korea}

\begin{abstract}
Changes of baryon properties in nuclear matter are investigated within 
the framework of an in-medium modified SU(3) Skyrme model.
Introducing the medium functionals in the SU(2) sector and
considering the alteration of kaon properties in nuclear medium,
we are able to examine the medium modification of the nucleon and
hyperons. The functionals introduced in the SU(2) sector are related
to ordinary nuclear matter properties near the saturation point. The
results indicate that the changes of the baryon properties in the
strange sector are strongly correlated with the in-medium properties
of kaons.
\keywords{Skyrmions, nucleons, hyperons, nuclear matter}
\end{abstract}

\maketitle              

\section{Introduction}
Understanding how the hyperons undergo changes in nuclear matter is a
very important issue in contemporary nuclear physics. In particular,
it is of great interest to see how the hyperons are related to
in-medium kaon properties at low densities and
how they can be changed in higher densities that can be found in the
interior 
of neutron stars.~\cite{Gal:2016boi,Lattimer:2015nhk}.
In the present contribution, we will discuss a recent work on the
hyperon properties in nuclear matter, which was carried out in a
simple but plausible framework of a chiral soliton approach to
nonzero density phenomena in the SU(3) sector~\cite{Hong:2018sqa}.
Previously, a similar approach was developed in the non-strangeness
sector to study various phenomena in medium (for example, see
Ref.~\cite{Yakhshiev:2013eya}  and references therein) and the results
were in qualitative agreement with those from other different
approaches. In Ref.~\cite{Hong:2018sqa}, we extended the work of
Ref.~\cite{Yakhshiev:2013eya} to the SU(3) including the hyperons.
We discuss the main results and significance of the work.

\section{The model}
The Lagrangian of the present model written by the following
form~\cite{Hong:2018sqa}
\begin{eqnarray}
  \mathcal{L}=&-\frac{F_\pi^2}{16}\alpha_2^t(\rho)
  {\rm Tr} L_0L_0+\frac{F_\pi^2}{16}\alpha_2^s(\rho){\rm Tr} L_iL_i
  -\frac{\alpha_4^t(\rho)}{16e^2}
  {\rm Tr}[L_0,L_i]^2 +\frac{\alpha_4^s(\rho)}{32e^2}{\rm Tr}[L_i,L_j]^2\cr
&  +\frac{F_\pi^2}{16}\alpha_{\chi SB}(\rho){\rm Tr}
  \mathcal{M}(U+U^\dagger-2)+ \mathcal{L}_{WZ},
\label{ModLag}
\end{eqnarray}
where $L_\mu=U^\dagger\partial_\mu U$ and $U(\bm{x},t)$ is a chiral
field in SU(3). The Wess-Zumino term~\cite{Wess:1971yu}
$\mathcal{L}_{\mathrm{WZ}}$ in the Lagrangian constrains the soliton
to be identified as a baryon and is expressed by a five-dimensional
integral over a disk $D$
\begin{eqnarray}
S_{\rm WZ} = -\frac{iN_c}{240\pi^2} \int_{D} d^5 \vec x\,
   \epsilon^{\mu\nu\alpha\beta\gamma}
   {\rm Tr}(L_\mu L_\nu L_\alpha L_\beta L_\gamma).
\end{eqnarray}
Here $\epsilon^{\mu\nu\alpha\beta\gamma}$ is the totally antisymmetric
tensor defined as $\epsilon^{01234}=1$ and $N_c=3$ is the number of
colors. The values of input parameters are defined in free space:
$F_\pi=108.783$\,MeV denotes the pion decay constant,  $e=4.854$
represents the Skyrme parameter, the masses of the $\pi$ and $K$
mesons are given respectively as $m_\pi=134.976$\,MeV and
$m_K=495$\,MeV, and the mass matrix of the pseudo-Nambu-Goldstone
bosons $\mathcal{M}$ has the diagonal form
$\mathcal{M}=(m_\pi^2,m_\pi^2,m_K^2)$.  The density-dependent
functions $\alpha_2^t(\rho)$, $\alpha_2^s(\rho)$, $\alpha_4^t(\rho)$,
$\alpha_4^s(\rho)$ and $\alpha_{\chi SB}(\rho)$ reflect the changes
of the meson properties in nuclear medium. In an approximation of
homogeneous infinite nuclear matter they are expressed in terms
of the three linear density-dependent functions
$f_{i}(\rho)=1+C_i\rho,\,(i=1,2,3)$. The numerical values of $C_i$ are
fixed to be $C_1=-0.279$,  $C_2=0.737 $ and $C_3=1.782$, respectively.
They reproduce very well the equations of state (EoS)
for symmetric nuclear  matter near the normal nuclear matter density
$\rho_0$ and
at higher densities that may exist in the interior of a neutron star.
The medium modification of the kaon properties is achieved by
considering the following scheme
\begin{eqnarray}
F_\pi m_K\rightarrow F_K^* m_K^*=F_\pi m_K(1-C\rho/\rho_0)
\label{comKprop}
\end{eqnarray}
and can be explained in terms of the alteration of the kaon decay
constant and/or of the kaon mass in nuclear environment.

The quantization of the model is performed by considering the
time-dependent rigid rotation of a static soliton
\begin{equation}
U(\bm{r},t)=\mathcal{A}(t)U_0(\bm{r})\mathcal{A}(t)^\dagger,
\end{equation}
where $U_0(\bm{r})$ denotes the static SU(3) chiral soliton with
trivial embedding. The time-dependent rotational matrix
$\mathcal{A}(t)$ is decomposed
\begin{eqnarray}
\mathcal{A}(t)&=\left(\begin{array}{cc}
A(t)&0\\
0^\dagger&1\end{array}\right)S(t),
\end{eqnarray}
in terms of the SU(2) isospin rotation
$A(t)=k_0(t){\bf 1}+i \sum_{a=1}^3\tau_a k_a(t)$ and
fluctuations into the strangeness sector given by the
matrix $S(t)=\exp\left\{i\sum_{p=4}^7k_p \lambda_p\right\}$.
Here $\tau_{1,2,3}$ denote the Pauli matrices, whereas $\lambda_p$
stand for the strange part of the SU(3) Gell-Mann matrices.
The time-dependent functions $k_a(t)$ $(a=0,1,2,\dots,7)$ represent
arbitrary collective coordinates.  The more details of the approach
can be found  in Ref.~\cite{Hong:2018sqa}.

\section{Results and discussions}
All model parameters in free space and in nuclear matter, except for
the parameter $C$ in Eq.~(\ref{comKprop}), are fixed
in the SU(2) sector. The only remaining parameter $C$ could be fixed
by data on kaon-nucleus scattering and kaonic atoms. However, in the
present work we carry out a qualitative analysis of the
effects in the baryonic sector due to the modification of the kaon
properties in nuclear medium. Consequently, we discuss the density
dependence of the mass splittings among the various baryon multiplet
members. In our calculation, the parameter value $C=0$ corresponds to
the case when the properties of kaon will not change in nuclear matter
whereas a nonzero value of the parameter $C\neq 0$ indicates that the
mass and/or kaon dynamics is alters in a dense nuclear environment.

The results show that in general the masses of the baryon
octet tend to decrease in nuclear matter. Only $\Sigma$
showed a different tendency if the parameter value is set to be $C=0$.
In the case of $C=0.2$, $m_\Sigma$ also tends to decrease  as
the density of nuclear matter increases~\cite{Hong:2018sqa}.
In comparison, the results from SU(3) chiral effective field
theory~\cite{Petschauer:2015nea} show that $m_{\Lambda}^*$ is
decreased by about 17~\% at normal nuclear matter density
$\rho_0$. The $\Xi$ hyperon is
behaved in a similar manner. At $\rho_0$  the change in the mass of
$\Xi^*$ was about 6~\% and 16~\% for the corresponding parameter
values $C=0$ and $C=0.2$, respectively.
The masses of the baryon decuplet increase in general as $\rho$
increases. Changes are dramatic for $C=0$ while for $C=0.2$ they are
less changeable.

\begin{figure}[th]
\includegraphics[width=0.45\textwidth]{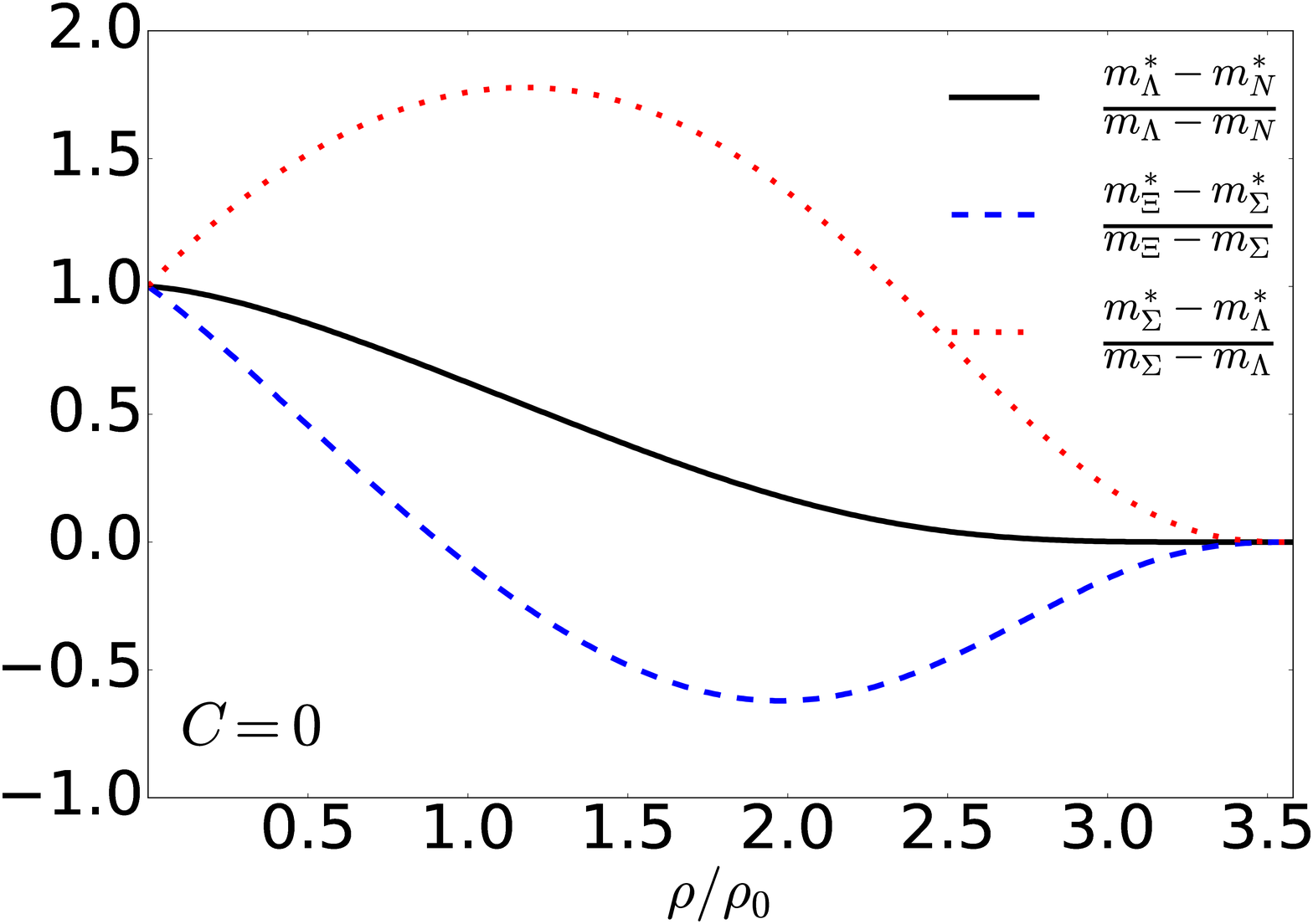}
\includegraphics[width=0.45\textwidth]{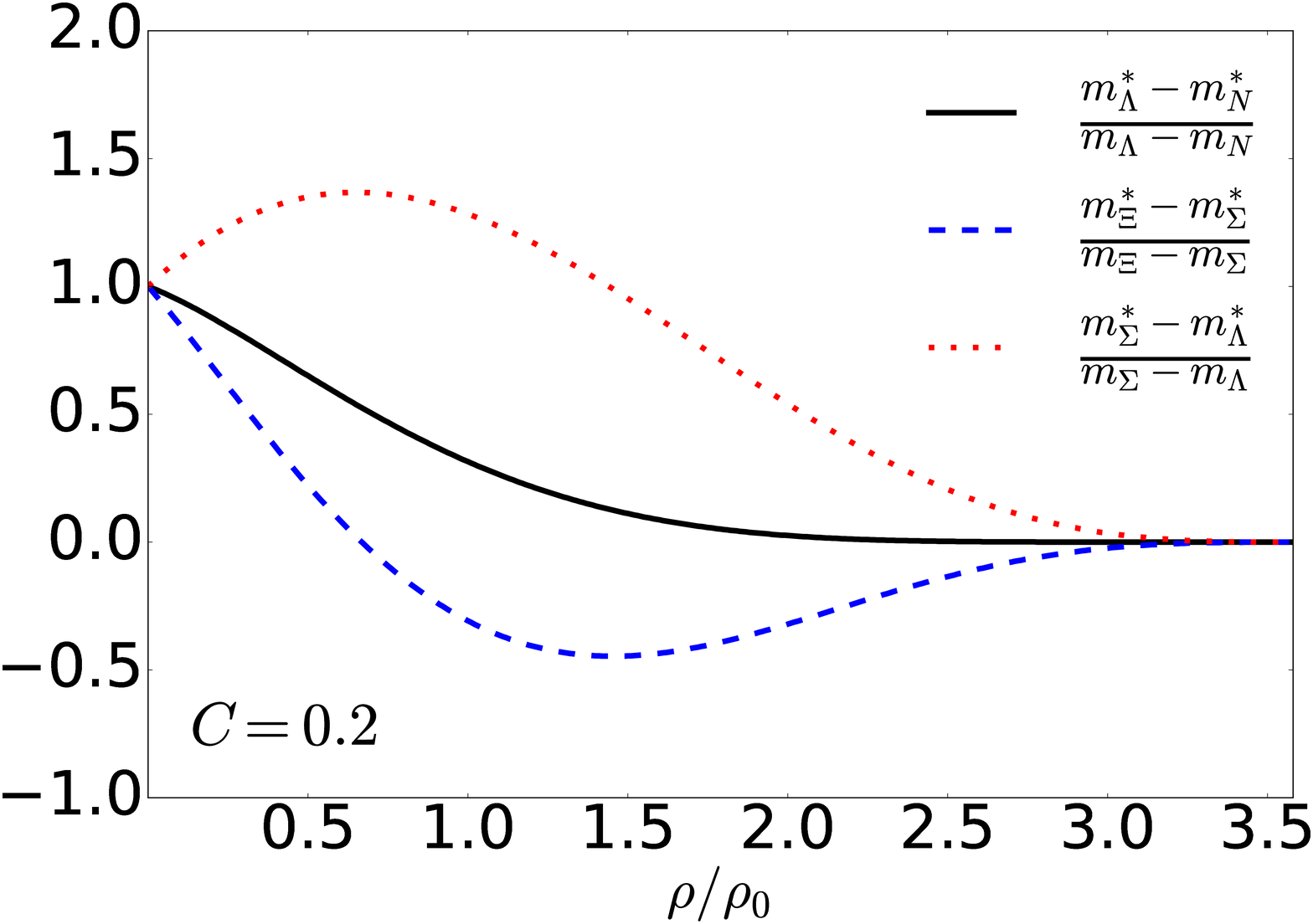}
\caption{ (Color online.) Density dependence of the mass splittings
among the baryon octet members. The mass splittings in nuclear matter
are normalized to the corresponding free space mass
splittings. The left and right panels in the figure corresponds to the
results with $C=0$ and $C=0.2$, respectively.
}
\label{Fig1}
\end{figure}

\begin{figure}[th]
\includegraphics[width=0.45\textwidth]{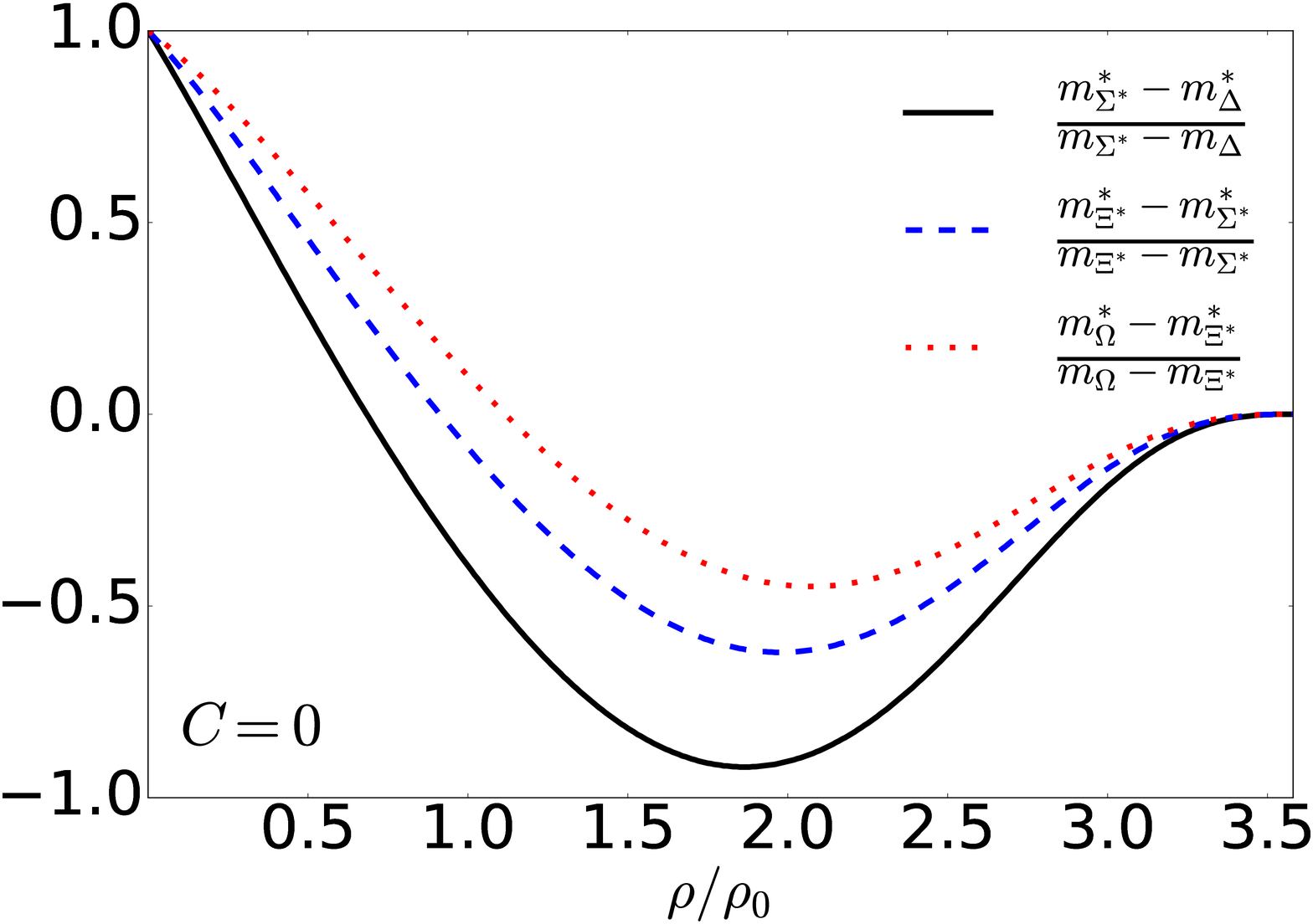}
\includegraphics[width=0.45\textwidth]{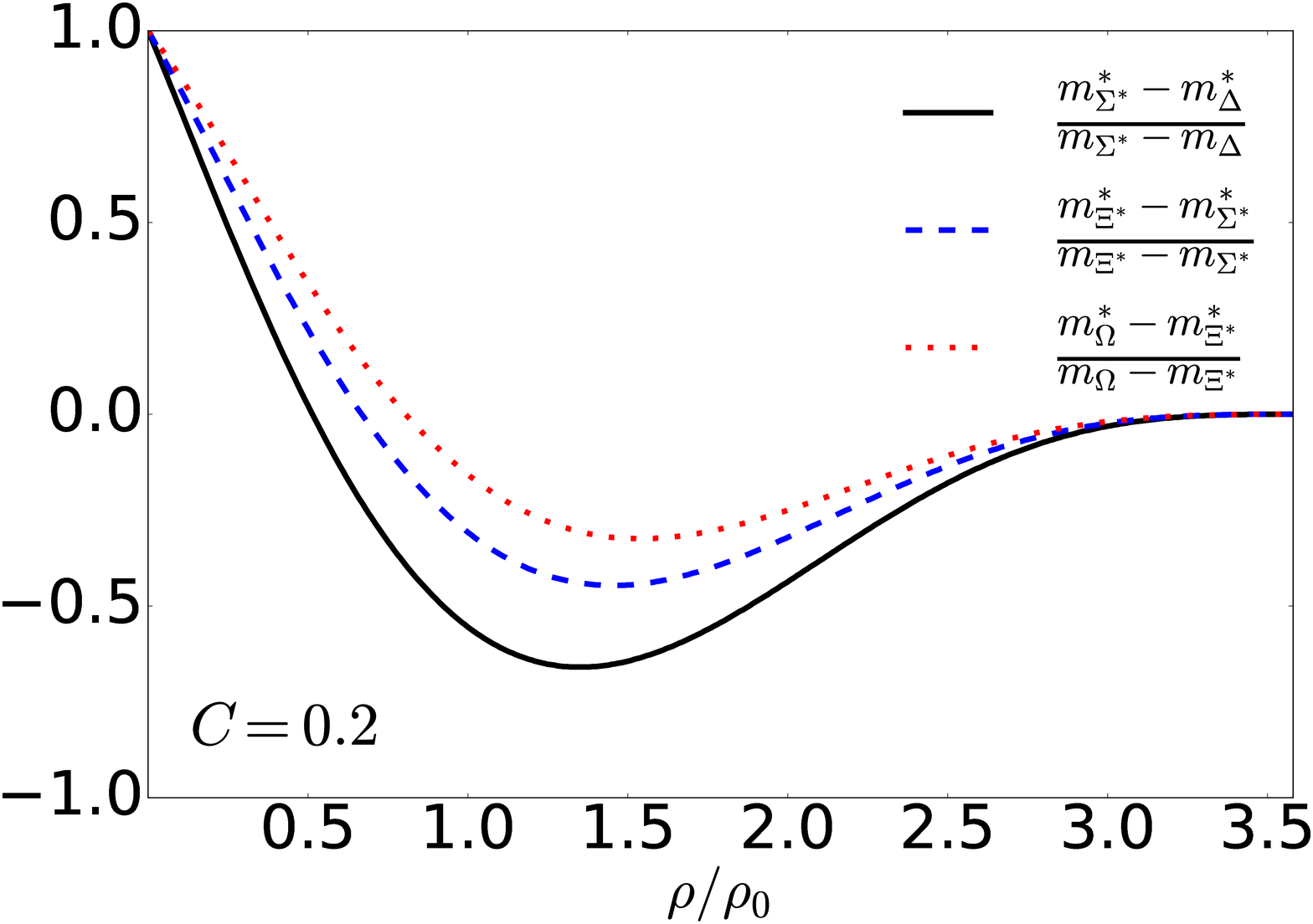}
\caption{ (Color online.) Density dependence of the mass splittings
among the baryon decuplet members.
Notations are the same as in Fig.~\ref{Fig1}.
}
\label{Fig2}
\end{figure}

We present the density dependence of the mass splittings among the
multiplet members in Figs.~\ref{Fig1} and \ref{Fig2}.
Figure~\ref{Fig1} shows the density dependence of the mass splittings
among the baryon octet members while Fig.~\ref{Fig2} depicts the results
corresponding to the mass splittings among the decuplet members.
All the mass splittings in nuclear matter are normalized
to the vaues of the corresponding ones in free space.
The left and right panels in the figures illustrate the results with
two different values of parameter $C$, respectively.

It is interesting to see that except $m^*_{\Sigma}-m^*_{\Lambda}$
all the mass splittings tend to decrease
up to $(1.5-2)\rho_0$. This behavior can be explained
in terms of the density-dependent functionals $\omega^*_-$ and $c^*$
entering into the mass formula (see Eq.\,(36) in
Ref.~\cite{Hong:2018sqa}).  The first functional describes the
fluctuations in the strangeness direction and comes into play for the
mass splitting formula between the same strangeness members while all
other mass splittings presented in the figures depend linearly on
$\omega_-^*$.
This indicates that at large densities the fluctuations in strangeness
direction gets weaker. From the figures one concludes also that
at large densities SU(3) flavor symmetry tends to be restored.

The work is supported by Basic Science Research Program through the
National Research Foundation (NRF) of Korea funded by the
Korean government (Ministry of Education, Science and Technology,
MEST), Grant No. 2016R1D1A1B03935053 (UY) and Grant
No. NRF-2018R1A2B2001752 (HChK).

\end{document}